\documentclass[aps,preprint,prd,nofootinbib]{revtex4}
\usepackage{graphicx}
\usepackage{psfrag}

\usepackage{subfigure}
\usepackage{array}
\usepackage{graphicx}
\usepackage{amsmath}      
\usepackage{amssymb}
\usepackage{mathrsfs}       
\usepackage{bm}            
\usepackage{slashed}
\usepackage{threeparttable}
\usepackage{multirow}
\usepackage{appendix}
\usepackage[colorlinks, linkcolor=black, anchorcolor=black, citecolor=black]{hyperref}
\usepackage[amsmath,thmmarks,hyperref]{ntheorem}
\usepackage{soul}

\begin{document}

\title{\fontsize{19pt}{19pt}Strong decays of $D_{3}^{\ast}(2760)$, $D_{s3}^{\ast}(2860)$, $B_{3}^{\ast}$, and $B_{s3}^{\ast}$}
\author{Tianhong Wang$^1$\footnote{thwang@hit.edu.cn}, Zhi-Hui Wang$^2$\footnote{wzh19830606@163.com}, Yue Jiang$^1$\footnote{jiangure@hit.edu.cn}, Libo Jiang$^3$\footnote{jiangl@fnal.gov}\\and Guo-Li Wang$^1$\footnote{gl\_wang@hit.edu.cn}\\}
\address{$^1$Department of Physics, Harbin Institute of Technology,
Harbin, 150001, China\\
$^2$School of Electrical $\&$ Information Engineering, Beifang University of Nationalities, Yinchuan, 750021, China\\
$^3$Department of Physics and Astronomy, University of Pittsburgh, Pittsburgh, Pennsylvania 15260, USA}

\baselineskip=20pt

\begin{abstract}
In this paper, we study the OZI-allowed two-body strong decays of $3^-$ heavy-light mesons. Experimentally the charmed $D_{3}^{\ast}(2760)$ and the charm-strange $D_{s3}^{\ast}(2860)$ states with these quantum numbers have been discovered. For the bottomed $B(5970)$ state, which was found by the CDF Collaboration recently, its quantum number has not been decided yet and we assume its a $3^-$ meson in this paper. The theoretical prediction for the strong decays of bottom-strange state $B_{s3}^\ast$ is also given. The relativistic wave functions of $3^-$ heavy mesons are constructed and their numerical values are obtained by solving the corresponding Bethe-Salpeter equation with instantaneous approximation. The transition matrix is calculated by using the PCAC and low energy theorem, following which, the decay widths are obtained. For $D_{3}^\ast(2760)$ and $D_{s3}^\ast(2860)$, the total strong decay widths are 72.6 MeV and 47.6 MeV, respectively. For $B_3^\ast$ with $M=5978$ MeV and $B_{s3}^\ast$ with $M=6178$ MeV, their strong decay widths are 22.9 MeV and 40.8 MeV, respectively.
\end{abstract}

\maketitle

\section{Introduction}

In the last few years, many new hadron states have been discovered experimentally, injecting new vitality to the study of hadron physics. Among these new states, some are thought to be tetraquark, pentaquark~\cite{pentaquark}, or molecule states, while some are believed to have the usual quark-antiquark structure~\cite{opencharm}. The observation of the second case improves the meson spectra predicted by the quark potential models and may bring more insights into the nonperturbative properties of QCD. Among these particles, we are interested in the spin-3 heavy-light mesons in this paper, as more data about such states are collected recently. In 2006, the Babar Collaboration found the $D_{sJ}^\ast(2860)$ state~\cite{babar1} which was confirmed by LHCb~\cite{LHCb}. This particle attracted much attention~\cite{Close1, Col3, Eef, Guo, ZZ08, God01, Vi09, Seg1, Song01, ZGW15, ZGW16, ZGW17, Bad, God16}. Theoretically it is thought to be a charm-strange meson with spin-parity quantum number $J^P=3^-$ or $1^-$ (S-D mixing). Both predict the correct partial decay widths within the experimental error.  This uncertainty was eliminated in 2014 by the LHCb Collaboration~\cite{Aa01, Aa02} which found that two particles, namely, $D_{s3}^\ast(2860)$ with spin-3 and $D_{s1}^\ast(2860)$ with spin-1, are around this mass region.

For the charmed meson, $D^\ast(2760)$ was discovered by the BaBar Collaboration~\cite{babar3} and $D_J^\ast(2760)$ was found by LHCb~\cite{LHCb13}. Both particles have similar masses and decay widths, so they are thought to be the same state. Just as $D_{sJ}^\ast(2860)$, they are also thought to be $3^-$ or $1^-$ state. Recently, LHCb~\cite{LHCb15} found the first spin-3 charmed meson $D_3^\ast(2760)$, whose decay width (of the Isobar formalism, see Table III) is about 30 MeV larger than that of $D_J^\ast(2760)$~\cite{LHCb13}. Whether a $1^-$ partner with the similar mass with $D_3^\ast(2760)$ exists (as the charm-strange case) is an interesting question. In the bottomed (bottom-strange) meson sector, the $3^-$ state has not been found. However, very recently the CDF Collaboration reported the existence of $B(5970)$~\cite{CDF}, which has been investigated by assuming it has the quantum number $1^-$~\cite{Sun14, ZGW14} or $3^-$~\cite{ZGW14}. The decay width still has large experimental error (see Table I), so more precise detection is needed.

Usually, if the strong decay channels of a meson are OZI-allowed, they will be dominant, and the sum of their partial widths can be used to estimate the total width of the meson. Beside that, those decays are also applied to determine the quantum number of particles. To study these decays, several theoretical methods could be applied, such as the chiral quark method~\cite{ZZ08, ZZ, Zhong10, Xiao14, Mat02}, the heavy meson effective theory~\cite{Col2, Col3, ZGW14, ZGW15}, the QCD sum rules~\cite{ZGW16, ZGW17}, and the $^3P_0$ method~\cite{God01,Li, Li11, Zhang, God14, Song02,Sun14, Seg1,Chen15,God16, GMS, Lu, Lu14}.  The chiral quark model introduces an effective Lagrangian to describe the coupling between light quark fields and light meson, while for the heavy meson effective theory, the interaction lagrangian is constructed just by meson fields. The $^3P_0$ model is very popular in dealing with OZI-allowed strong decays. In this method, a $q\bar q$ with $J^{PC}=0^{++}$ is assumed to be created from the vacuum. For the heavy mesons, the simple harmonic oscillator (SHO) wave functions are usually adopted. 

In our recent work~\cite{QL02}, the weak production of $3^-$ heavy-light states from the $D(D_s)$ or $B(B_s,B_c)$ mesons have been studied. When these particles are produced, they will decay very quickly to the lighter final states which are used experimentally to reconstruct their mother particle.  Here, by using the same formalism, we investigate the OZI-allowed two-body strong decays of these $3^-$ mesons, This may be helpful to gain more information of these high-spin states, especially for the undiscovered $b$-flavored ones.

As Figure 1 shows, the OZI-allowed two body strong decays can be realized by introducing a scalar type interaction vertex. It can also be realized without that interaction vertex, that is the light quark and antiquark are connected by a propagator, which is used in Ref.~\cite{Ric} and our previous work~\cite{OZI13}. Under the current situation, there is a light meson in the final states, whose wave function cannot be described by the instantaneous approximation. So to deal with this difficulty, we take a different method, which is realized by using the reduction formula, PCAC and the low energy theorem. This method has been applied to deal with the strong decays of S-wave heavy-light mesons~\cite{ZHW, ZHW1}, which get the results close to the experimental data. However, this method can only be applied to the case when the light meson being a pseuscalar one. For the case when the light meson is vector, PCAC cannot be used. For those channels, we will adopt an effective lagrangian to describe the quark-meson coupling.

\begin{figure}[ht]\label{Feyn2}
\centering
\includegraphics[scale=1.05]{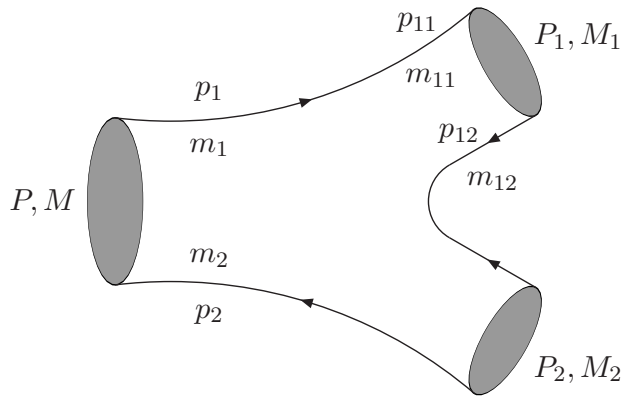}
\caption[]{Feynman diagram of the OZI-alllowed two-body strong decay channel.}
\end{figure}

Since the relativistic effects should be considered, especially for the state with high orbital angular momentum,  using more appropriate wave functions to calculate the strong decays of these high-spin mesons is necessary. In this paper, the instantaneous Bethe-Salpeter equation~\cite{BS1, BS2}, namely, the full Salpeter equation is used to get the mass spectrum and corresponding wave functions of heavy-light mesons. The transition matrix can be written within Mandelstam formalism~\cite{Man}.   

The paper is organized as follows. In Section II, we present the theoretical formalism of the calculation. The wave function of the $3^-$ state is constructed. For the channels with a light pseudoscalar meson, the quark-meson coupling is introduced by two methods, while for the light vector case, an effective Lagrangian from other literature is adopted. In Section III, we give the results of strong decays of four heavy-light mesons and compare them with those of other models. Finally, we draw the conclusion in Section IV. 

\section{Theoretical formalism}

As the wave functions of heavy mesons will be used in the following to calculate the transition amplitude, it must be constructed as a starting point. In our previous works~\cite{WTH01, QL01, QL02}, the wave function of the $3^{-}$ state has been given as
\begin{equation}
\begin{aligned}
\varphi_{3^{-}}(q_\perp)& = \epsilon_{\mu\nu\alpha}q_\perp^\mu q_\perp^\nu\Big\{q_\perp^\alpha\Big(f_1 + \frac{\slashed P}{M} f_2 + \frac{\slashed q_\perp}{M}f_3 + \frac{\slashed P\slashed q_\perp}{M^2} f_4\Big)+M\gamma^\alpha \Big(f_5 + \frac{\slashed P}{M} f_6 \\
&+ \frac{\slashed q_\perp}{M}f_7 + \frac{\slashed P\slashed q_\perp}{M^2} f_8\Big)\Big\},
\end{aligned}
\end{equation}
where $M$ and $P$ are the mass and momentum of the meson, respectively; $q$ is the relative momentum between the quark and antiquark; $q_\perp$ is defined as $q-\frac{P\cdot q}{M}P$; $f_i$s are functions of $q_\perp$ which will be obtained by solving the full Salpeter equation; $\epsilon_{\mu\nu\gamma}$ is the polarization tensor of the meson, which is totally symmetric and satisfies 
\begin{equation}
g^{\mu\nu}\epsilon_{\mu\nu\gamma}=0, \hspace{7mm} P^\mu\epsilon_{\mu\nu\gamma}=0.
\end{equation}
The completeness relation is given by~\cite{tensor}
\begin{equation}
\begin{aligned}
\sum_{\lambda=-3}^{3} \epsilon^{(\lambda)}_{abc}\epsilon^{\ast(\lambda)}_{xyz} &=\frac{1}{6}(\mathcal{P}_{ax}\mathcal{P}_{by}\mathcal{P}_{cz}+\mathcal{P}_{ax}\mathcal{P}_{bz}\mathcal{P}_{cy}+\mathcal{P}_{ay}\mathcal{P}_{bx}\mathcal{P}_{cz}
+\mathcal{P}_{ay}\mathcal{P}_{bz}\mathcal{P}_{cx}\\
&+\mathcal{P}_{az}\mathcal{P}_{by}\mathcal{P}_{cx}+\mathcal{P}_{az}\mathcal{P}_{bx}\mathcal{P}_{cy})
-\frac{1}{15}(\mathcal{P}_{ab}\mathcal{P}_{cz}\mathcal{P}_{xy}+\mathcal{P}_{ab}\mathcal{P}_{cy}\mathcal{P}_{xz}\\
&+\mathcal{P}_{ab}\mathcal{P}_{cx}\mathcal{P}_{yz}
+\mathcal{P}_{ac}\mathcal{P}_{bz}\mathcal{P}_{xy}+\mathcal{P}_{ac}\mathcal{P}_{by}\mathcal{P}_{xz}+\mathcal{P}_{ac}\mathcal{P}_{bx}\mathcal{P}_{yz}\\
&+\mathcal{P}_{bc}\mathcal{P}_{az}\mathcal{P}_{xy}+\mathcal{P}_{bc}\mathcal{P}_{ay}\mathcal{P}_{xz}+\mathcal{P}_{bc}\mathcal{P}_{ax}\mathcal{P}_{yz}),
\end{aligned}
\end{equation}
where we have defined $\mathcal{P}_{\mu\nu} \equiv -g_{\mu\nu} + \frac{P_\mu P_\nu}{M^2}$.

\begin{figure}[ht]\label{Feyn1}
\centering
\includegraphics[scale=1.05]{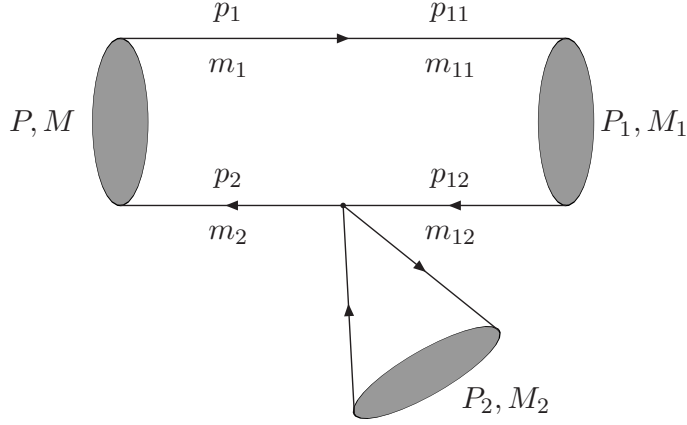}
\caption[]{Feynman diagram of the OZI-allowed two-body strong decay channel of the heavy-light meson with the interaction vertex being changed the form.}
\end{figure}

By using the reduction formula, the transition amplitude can be written as the production of the inverse propagator and the expectation value of the light meson field~\cite{chang}. We take the process $D_{sJ}^{\ast}\rightarrow D^{(\ast)}K$ as an example, which has the form
\begin{equation}
\begin{aligned}
\langle D^{(\ast)}(P_1)K(P_2)|D^{\ast+}_{sJ}(P)\rangle = \int d^4xe^{iP_2\cdot x}(M_K^2-P_2^2)\langle D^{(\ast)}(P_1)|\Phi_K(x)|D^{\ast+}_{sJ}(P)\rangle.
\end{aligned}
\end{equation}
By using PCAC, the light meson field is expressed as the divergence of the axial-vector current divided by the decay constant of the light meson
\begin{equation}
\Phi_K(x) = \frac{1}{M_K^2f_K}\partial^\mu(\bar q\gamma_\mu\gamma_5 s).
\end{equation}
Combining Eq.~(4) and Eq.~(5), we get
\begin{equation}
\begin{aligned}
\langle D^{(\ast)}(P_1)K(P_2)|D^{\ast+}_{sJ}(P)\rangle &= \frac{M_K^2-P_2^2}{M_K^2f_K} \int d^4xe^{iP_2\cdot x}\langle D^{(\ast)}(P_1)|\partial^\mu(\bar q\gamma_\mu\gamma_5 s)|D^{\ast+}_{sJ}(P)\rangle\\
&=\frac{-iP_2^\mu(M_K^2-P_2^2)}{M_K^2f_K} \int d^4xe^{iP_2\cdot x}\langle D^{(\ast)}(P_1)|\bar q\gamma_\mu\gamma_5 s|D^{\ast+}_{sJ}(P)\rangle,
\end{aligned}
\end{equation}
where in the second equation partial integral is used.
Finally, by using low-energy theorem~\cite{chang}, we can get the form of the transition amplitude in the momentum space (see Figure 2)
\begin{equation}
\begin{aligned}
\langle D^{(\ast)}(P_1)K(P_2)|D^{\ast+}_{sJ}(P)\rangle \approx (2\pi)^4\delta^4(P-P_1-P_2)\frac{-iP_2^\mu}{f_K}\langle D^{(\ast)}(P_1)|\bar q\gamma_\mu\gamma_5s|D^{\ast+}_{sJ}(P)\rangle.
\end{aligned}
\end{equation}

This result can also be achieved by adopting the effective lagrangian method~\cite{ZZ, Mat02},
\begin{equation}
\mathcal L_{qqP} = \frac{g}{\sqrt{2}f_h}\bar q_i\gamma_\mu\gamma_5q_j\partial^\mu\phi_{ij},
\end{equation}
where
\begin{equation}
\phi_{ij} = \sqrt{2}\left(
\begin{array}{ccc}
\frac{1}{\sqrt{2}}\pi^0 + \frac{1}{\sqrt{6}}\eta & \pi^+ & K^+\vspace{0.25cm}\\
\pi^- & -\frac{1}{\sqrt{2}}\pi^0 + \frac{1}{\sqrt{6}}\eta & K^0 \vspace{0.25cm}\\
K^- & \bar K^0 & -\frac{2}{\sqrt{6}}\eta\\
\end{array}
\right)
\end{equation}
is the chiral field of pseudoscalar mesons. The quark-meson coupling constant $g$ is taken to be unity. $f_h$ is the decay constant.

Within Mandelstam formalism, the transition amplitude can be written as the overlapping integral over the Salpeter wave functions of the initial and final mesons~\cite{chang}
\begin{equation}\label{M1}
\begin{aligned}
\mathcal M& = \frac{-iP_2^\mu}{f_K}\langle D^{(\ast)}(P_1)|\bar q\gamma_\mu\gamma_5s|D^{\ast+}_{sJ}(P)\rangle \\
&\approx \frac{-iP_2^\mu}{f_K}\int\frac{d^3\vec q}{(2\pi)^3}{\rm Tr}\left[\overline \varphi_{P_1}^{++}\left(\vec q - \frac{m_{1}^\prime}{m_{1}^\prime+m_{2}^\prime}\vec P_1\right)\frac{\slashed P}{M}\varphi_{P}^{++}(\vec q)\gamma_\mu\gamma_5\right],
\end{aligned}
\end{equation}
where $m_1^\prime$ and $m_2^\prime$ are respectively the masses of quark and antiquark in the final  $D^{(\ast)}$ meson; $\overline\varphi$ is defined as $\gamma^0\varphi\gamma^0$; $\varphi^{++}$ is the positive energy part of the wave function. In the above equation, we have neglected the contributions of negative energy part of the wave function, which is very small compared with that of the positive one (less than $1\%$).

If the final light meson is $\eta$ or $\eta^\prime$, we have to consider the $\eta-\eta^\prime$ mixing
\begin{equation}
\left(
\begin{array}{c}
\phi_\eta\vspace{0.15cm} \\ 
\phi_{\eta^\prime} \\ 
\end{array}
\right)=
\left(
\begin{array}{cc}
\cos\theta&  \sin\theta\vspace{0.15cm}\\
- \sin\theta & \cos\theta
\end{array}
\right)
\left(
\begin{array}{c}
\phi_{\eta_8}\vspace{0.15cm}\\
\phi_{\eta_0}
\end{array}
\right),
\end{equation}
where the mixing angle $\theta=19^\circ$ is used. The masses of physical states are related to the masses of flavor states by 
\begin{equation}
\left(
\begin{array}{c}
M^2_{\eta_8}\vspace{0.15cm} \\ 
M^2_{\eta_0} \\ 
\end{array}
\right)=
\left(
\begin{array}{cc}
\cos^2\theta&  \sin^2\theta\vspace{0.15cm}\\
\sin^2\theta & \cos^2\theta
\end{array}
\right)
\left(
\begin{array}{c}
M^2_{\eta}\vspace{0.15cm}\\
M^2_{\eta^\prime}
\end{array}
\right).
\end{equation}
By considering $\phi_{\eta_8}=(u\bar u + d\bar d -2s\bar s)/\sqrt{6}$ 
and $\phi_{\eta_0}=(u\bar u + d\bar d + s\bar s)/\sqrt{3}$, the transition amplitude of $D_{sJ}^\ast(2860)^+\rightarrow D_s^+\eta$ has the form~\cite{chang}
\begin{equation}
\begin{aligned}
\mathcal M = P_2^\mu\left[\frac{-2M_\eta^2\cos\theta}{\sqrt{6}M_{\eta_8}^2f_{\eta_8}} + \frac{M_\eta^2\sin\theta}{\sqrt{3}M_{\eta_0}^2f_{\eta_0}}\right]\langle D^{(\ast)}_s(P_1)|\bar s\gamma_\mu\gamma_5s|D^{\ast+}_{sJ}(P)\rangle,
\end{aligned}
\end{equation}
where $f_{\eta_0}$ and $f_{\eta_8}$ are the decay constants of $\eta_0$ and $\eta_8$, respectively.

The method above can only be applied to the processes when the light meson is a pseudoscalar. In the case when a light vector boson involves, we use the effective lagrangian method which is adopted in Ref.~\cite{ZZ}. The quark-meson coupling is described by the lagrangian
\begin{equation}
\begin{aligned}
\mathcal{L}_{qqV} = \sum_{j}\bar q_j(a\gamma_\mu + \frac{ib}{2m_j}\sigma_{\mu\nu}P_2^\nu)V^\mu q_j,
\end{aligned}
\end{equation}
where $V^\mu$ is the field of the light vector meson with momentum $P_2$; $a=-3.0$ and $b=2.0$ represent the vector and tensor coupling strength, respectively. In Ref.~\cite{ZZ}, this lagrangian is reduced to the nonrelativistic form and the harmonic oscillator wave functions are used. In our calculation, we use Eq.~(14) directly, and the full Salpeter wave functions are applied which could provide some comparison with the results in Ref.~\cite{ZZ}.

After finishing the trace and integral in Eq.~(\ref{M1}), we get the transition amplitudes which are expressed as several form factors
\begin{equation}
\begin{aligned}
\mathcal M_{D_{s3}^\ast(2860)\rightarrow DK} = \frac{-i}{f_K}\epsilon^{\alpha\beta\gamma}P_{1\alpha} P_{1\beta} P_{1\gamma} t_1,
\end{aligned}
\end{equation}
\begin{equation}
\begin{aligned}
\mathcal M_{D_{s3}^\ast(2860)\rightarrow D^\ast K} =\frac{-i}{f_K} \epsilon^{\alpha\beta\gamma}\epsilon^\delta_1\epsilon_{\alpha\delta\sigma\xi}P^\sigma P_{1}^{\xi}P_{1\beta}P_{1\gamma}t_2,
\end{aligned}
\end{equation}
\begin{equation}
\begin{aligned}
\mathcal M_{D_{s3}^\ast(2860)\rightarrow DK^\ast} =\epsilon^{\alpha\beta\gamma}\epsilon^\delta_2\epsilon_{\alpha\delta\sigma\xi}P^\sigma P_{1}^{\xi}P_{1\beta}P_{1\gamma}t_3,
\end{aligned}
\end{equation}
\begin{equation}
\begin{aligned}
\mathcal M_{D_3^\ast(2760)\rightarrow D_2^\ast(2460)\pi} = \frac{-i}{f_\pi}\epsilon^{\alpha\mu\sigma\delta}P_\sigma P_{1\delta}P_1^\beta \epsilon_{\alpha\beta\gamma}\epsilon_{1\mu\nu} (P_1^\gamma P^\nu t_4 + g^{\gamma\nu} M^2 t_5),
\end{aligned}
\end{equation}
\begin{equation}
\begin{aligned}
\mathcal M_{D_3^\ast(2760)\rightarrow D_1(2420)\pi} = \frac{-i}{f_\pi}\epsilon^{\alpha\beta\gamma}\epsilon^\mu_1 P_{1\beta} P_{1\gamma} (P_{1\alpha}  P_\mu t_6 + g_{\alpha\mu} M^2 t_7),
\end{aligned}
\end{equation}
\begin{equation}
\begin{aligned}
\mathcal M_{B_{3}^\ast\rightarrow B^\ast \rho} &=\epsilon^{\alpha\beta\gamma}P_{1\alpha}(P_{1\beta}\epsilon_1\cdot P\epsilon_{2\gamma}s_1+P_{1\beta}\epsilon_2\cdot P\epsilon_{1\gamma}s_2+P_{1\beta}P_{1\gamma}\epsilon_1\cdot P\epsilon_2\cdot P s_3/M^2 \\
&+P_{1\beta}P_{1\gamma}\epsilon_1\cdot\epsilon_2 s_4+ M^2 \epsilon_{1\beta}\epsilon_{2\gamma}s_5).
\end{aligned}
\end{equation}
In the above equations, $\epsilon^{\alpha\mu\sigma\delta}$ is the totally antisymmetric tensor; $\epsilon$, $\epsilon_1$, and $\epsilon_2$ are the polarization vectors (tensor) of the initial meson, the final heavy meson, and the final light meson, respectively. The form factors $t_1\sim t_7$ and $s_1\sim s_5$ are integrals of $q_\perp$. For different channels, the integrations have different expressions. Thus the form factors have different values. Here we just take some channels of $D_{s3}$, $D_3$ and $B_3^\ast$ as examples. Other decay channels would have the same form of form factors as one of above equations. Such as $\mathcal M_{D_3^\ast(2760)\rightarrow D_{1^\prime}(2420)\pi}$ would have the same expression as Eq.~(19).

The two-body decay width is
\begin{equation}
\begin{aligned}
\Gamma=\frac{|\vec P_1|}{8\pi M^2}\frac{1}{2J+1}\sum_\lambda|\mathcal M|^2,
\end{aligned}
\end{equation}
where $|\vec P_1| = \sqrt{[M-(M_1-M_2)^2][M-(M_1 + M_2)^2]}/2M$ is the momentum of the final meson; $J=3$ is the spin quantum number of the initial meson; $\lambda$ represents the polarization of both initial and final mesons.

\section{Results and discussions}

\begin{figure}
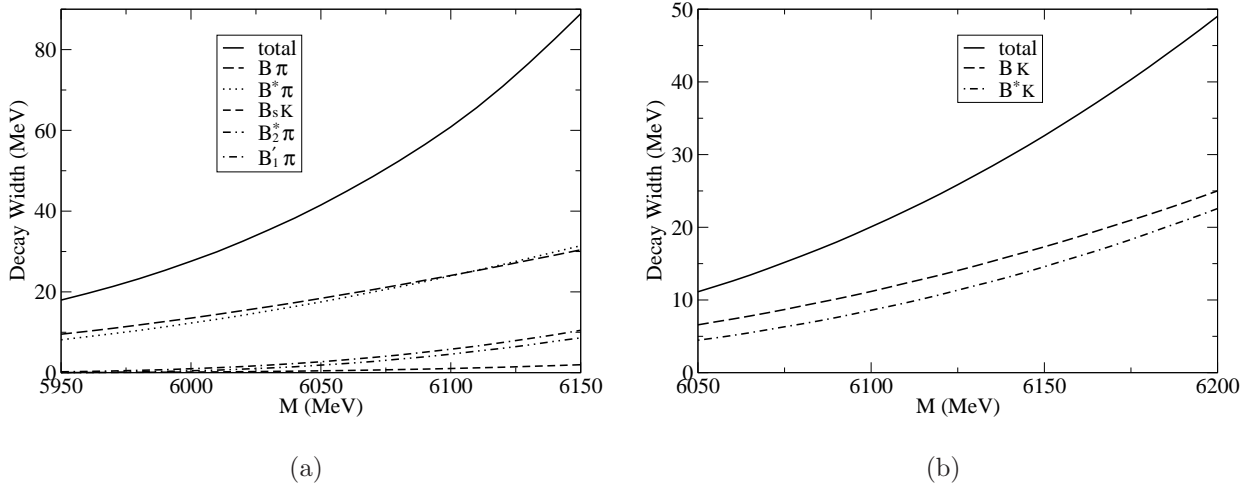

\centering
\subfigure[]{\includegraphics[scale=0.32]{B3.eps}}
\hspace{3mm}
\subfigure[]{\includegraphics[scale=0.32]{Bs3.eps}}
\caption[]{Two-body strong decay widths change with the mass of $3^-$ bottom and bottom-strange mesons. Only the dominant channels are considered. (a) is for $B_3^\ast$ and (b) is for $B_{s3}^\ast$.}
\end{figure}

To get the wave functions of the initial and final heavy mesons, we solve the full Salpeter equaiton. The interaction potential can be phenomenologically written as the Coulomb-like term (comes from one-gluon exchange) plus a linear term. We will not present the explicit form of the equation which can be found in Refs.~\cite{Kim, WTH01}. Here we just list the parameters used in the calculation: $m_u=0.305$ GeV, $m_d=0.311$ GeV, $m_s=0.5$ GeV, $m_c=1.62$ GeV, and $m_b=4.96$ GeV. For the masses of $D_{s3}^\ast$ and $D_3^\ast$, we will use the experimental data as the input value. For $B_3^\ast$, we will study two cases: $M=5978$ MeV (to compare with experimental result) and $M=6015$ MeV (to compare with the results of other models). As to $B_{s3}^\ast$ meson, we will use 6178 MeV to compare with Refs.~\cite{Sun14, GMS}. When the transition amplitude is calculated, the following parameters are adopted: $f_\pi$= 130.4 MeV, $f_K$= $156.2$ MeV~\cite{PDG}, $f_{\eta_8} = 1.26 f_\pi$, $f_{\eta_0} = 1.07 f_\pi$, $M_{\eta_8} =604.7 $ MeV, and $M_{\eta_0} = 923.0$ MeV~\cite{chang}.

The decay widths for $D_{s3}^\ast$ calculated by different models are listed in Table II. The dominant channels are $DK$ and $D^\ast K$, which in our calculation have partial widths 31.1 MeV and 14.6 MeV, respectively. Here we use $D^{(\ast)}K$ to represent $D^{(\ast)+}K^0+D^{(\ast)0}K^+$. Our results are close to those of other models, except that $\Gamma_{D^\ast K}$ in Ref.~\cite{Li} is about two times of ours. Refs.~\cite{Zhang, Li, Chen15, God14, Song02} use the $^3P_0$ model but with different parameter values, which causes diverse results. The chiral quark model is applied in Ref.~\cite{ZZ}. There for heavy mesons, the SHO wave functions is adopted. One can see that their results are smaller than ours. For the $DK^\ast$ channel, we use the same effective lagrangian form with that in Ref.~\cite{ZZ}, whose result is about two times smaller than ours. The total decay width for our model is close to the central value of the LHCb's result~\cite{Aa01, Aa02}, which is also at the same order with those of other models.  

For the $D_3^\ast$, the results of different models are presented in Table III. In our calculation, the partial widths of two dominant channels $D\pi$ and $D^\ast\pi$ are respectively 33.1 MeV and 22.0 MeV, which are consistent with those of other models, especially the chiral quark model~\cite{Zhong10}. For the channels with light vector meson $D\rho$ and $D\omega$, our results are about 4 times of those in Ref~\cite{Zhong10}, but compatible with those of the $^3P_0$ model~\cite{Li11}. Ref.~\cite{Yu15} also uses the $^3P_0$ model, but they get very large widths for these two channels, which makes the total width larger. In Table III, the decay width of $D_J^\ast(2760)$~\cite{LHCb13} is very close to our result, while for $D_3^\ast(2760)$~\cite{LHCb15}, as we pointed out before, its width is 30 MeV larger. Both results have large errors, which need more experimental observation.

In Table IV, the decay widths for $B_3^\ast$ is given. To compare with the results of other models, we consider two cases with different mass of $B_3^\ast$. For $M=5978$ MeV, the total decay width (22.9 MeV) is about 3 times smaller than the central value of the experimental data ($70^{+30}_{-20}\pm 30$ MeV) which has large errors. So we expect more data about this particle will be accumulated and more precise decay widths will be given. In Ref.~\cite{ZGW14}, the effective theory is used. There the experimental value is used to deduce the effective coupling which is applied to calculate the partial decay widths, the first two of which are about 3 times as large as ours. Ref.~\cite{Xiao14} gets the total decay width of $60$ MeV, which is 2 times larger than ours. When $M$ is taken to be $6105$ MeV, our results increase by about two times, which is about 2 and 4 times of those in Ref.~\cite{GMS} and Ref.~\cite{Sun14}. In our calculation, the $B^\ast_2\pi$ and $B_1^\prime\pi$ channels also give sizable contribution, which may be detected in the future to clarify the properties of this particle. For the mass of $B_2^\ast$, we take the value in PDG~\cite{PDG}, which is 50 MeV smaller than that taken in Ref.~\cite{Sun14} and Ref.~\cite{GMS}. Both references use the $^3P_0$ method and SHO wave functions. In Figure 3(a), we plot the total and main partial decay widths of $B_3^\ast$, where $M_{B_3^\ast}$ is taken to be 5950 MeV $\sim 6150$ MeV. The total width changes from 18 MeV to 89 MeV, which implies it depends strongly on the mass. One also notices that with the increase of mass, the decay width increases more and more quickly. 

The results for $B_{s3}^\ast$ is given in Table V. $BK$ and $B^\ast K$ give the main contribution. For the total decay width, we get 40.8 MeV which is larger than those in Ref.~\cite{Sun14} and Ref.~\cite{GMS} but smaller than that in Ref.~\cite{Lu}, where $^3P_0$ model is applied. Ref.~\cite{Xiao14} uses the chiral quark model. One can see a result about 2 times of ours is achieved when $M$ takes the same value.  Figure 3(b) shows when $M$ changes from 6050 MeV to 6200 MeV, the total decay width increases from 11 MeV to 49 MeV. As LHCb running, we expect this state will be detected in the near future.

An experimentally measured quantity is the ratio of the partial widths of two dominant decay channels. In Table VI, we present both theoretical and experimental results for this quantity of four heavy-light mesons.  For $\Gamma[D_{s3}^\ast\rightarrow D^\ast K]/\Gamma[D_{s3}^\ast\rightarrow DK]$, the experimental value (for $D_{sJ}^\ast(2860)$) is around 1, which is about two times of the theoretical predictions. In Ref.~\cite{ZZ}, a two-state scenario ($1{^3D_3}$ and $1D_{2^\prime}$) is proposed to explain this deviation. As LHCb has found there are two states $D_{s1}^\ast$ and $D_{s3}^\ast$ around 2860 MeV, more precise measurement of this ratio is needed. The ratio $\Gamma[D_{3}^\ast\rightarrow D^\ast \pi]/\Gamma[D_{3}^\ast\rightarrow D\pi]$ is close to that of the $D_{s3}^\ast$ case as a result of the $SU(3)_F$ symmetry. Our result is close to those of Refs.~\cite{Chen15, Xiao14}. For $\Gamma[B_{3}^\ast\rightarrow B^{\ast}\pi]/\Gamma[B_{3}^\ast\rightarrow B\pi]$, we present two results, which correspond $M_{B_{s3}^\ast}$= 6105 MeV and 5978 MeV (in the parenthesis), respectively. One can see our result is close to those of Refs.~\cite{GMS, Xiao14}. When $M_{B_3^\ast}$ takes $5950\sim6150$ MeV, this ratio changes from 0.85 to 1.03. The ratio $\Gamma[B_{s3}^\ast\rightarrow B^{\ast}\pi]/\Gamma[B_{s3}^\ast\rightarrow B\pi]$ is close to that of the $B_3^\ast$ case, which changes from 0.68 to 0.90 when $M_{B_{s3}^\ast}$ takes $6050\sim 6200$ MeV. 

\begin{table}
\caption{The experimental results of the mass (MeV) and decay width (MeV) for the candidates of heavy-light states with quantum number $3^-$.}
\vspace{0.2cm}
\setlength{\tabcolsep}{0.01cm}
\centering
\begin{tabular*}{\textwidth}{@{}@{\extracolsep{\fill}}cccc}
\hline\hline
State&Mass (MeV)&Width (MeV)&Reference\\ \hline
{\phantom{\Large{l}}}\raisebox{+.2cm}{\phantom{\Large{j}}}
$D_{sJ}^\ast(2860)$&$2856 \pm 1.5 \pm 5.0$&$47 \pm 7 \pm10$&BaBar~\cite{babar1}\\
{\phantom{\Large{l}}}\raisebox{+.2cm}{\phantom{\Large{j}}}
&$2866.1 \pm 1.0 \pm 6.3$&$69.9 \pm 3.2 \pm 6.6$&LHCb\cite{LHCb}\\
{\phantom{\Large{l}}}\raisebox{+.2cm}{\phantom{\Large{j}}}
&$2862\pm2^{+5}_{-2}$&$48 \pm 3 \pm 6$&BaBar~\cite{babar2}\\
{\phantom{\Large{l}}}\raisebox{+.2cm}{\phantom{\Large{j}}}
$D_{s3}^\ast(2860)$&$2860.5\pm2.6\pm2.5\pm6.0$&$53\pm7\pm4\pm6$&LHCb\cite{Aa01, Aa02}\\
{\phantom{\Large{l}}}\raisebox{+.2cm}{\phantom{\Large{j}}}
$D^\ast(2760)$&$2763.3\pm 2.3\pm2.3$& $60.9\pm5.1\pm3.6$&BaBar\cite{babar3}\\
{\phantom{\Large{l}}}\raisebox{+.2cm}{\phantom{\Large{j}}}
$D_J^\ast(2760)$&$2761.1\pm5.1\pm6.5 $&$74.4\pm3.4\pm37.0$&LHCb\cite{LHCb13}\\
{\phantom{\Large{l}}}\raisebox{+.2cm}{\phantom{\Large{j}}}
$D_{3}^\ast(2760)$&$2798 \pm 7 \pm 1 \pm 7$&$105 \pm 18 \pm 6 \pm 23$&LHCb\cite{LHCb15}\\
{\phantom{\Large{l}}}\raisebox{+.2cm}{\phantom{\Large{j}}}
$B(5970)$&$5978\pm 5\pm 12$&$70^{+30}_{-20}\pm 30$&CDF\cite{CDF}\\
\hline\hline
\end{tabular*}
\end{table}

\begin{table}
\caption{ Two-body strong decay widths (MeV) of $D_{s3}^{\ast}(2860)$. $^3P_0$ model is adopted in Refs.~\cite{Chen15}\cite{Li}\cite{Zhang}\cite{God14}\cite{Song02} and the chiral quark model is adopted by Ref.~\cite{ZZ}.}
\vspace{0.2cm}
\setlength{\tabcolsep}{0.2cm}
\centering
\begin{tabular*}{\textwidth}{@{}@{\extracolsep{\fill}}cccccccc}
\hline\hline
Mode&Ours&Ref.~\cite{Li}&Ref.~\cite{Zhang}&Ref.~\cite{Chen15}&Ref.~\cite{God14}&Ref.~\cite{ZZ}&Ref.~\cite{Song02}\\
\hline
{\phantom{\Large{l}}}\raisebox{+.2cm}{\phantom{\Large{j}}}
$DK$&31.1&35.6&22&28.5&20&24.1&$25\sim 30$ \\ 
{\phantom{\Large{l}}}\raisebox{+.2cm}{\phantom{\Large{j}}}
$D^{\ast}K$&14.6 &26.8&13&12.2&12&9.7&$14\sim24$\\
{\phantom{\Large{l}}}\raisebox{+.2cm}{\phantom{\Large{j}}}
$D_s\eta$&1.12 &1.6&1.2&1.9&1.0&1.7&$\sim 0.1$\\ 
{\phantom{\Large{l}}}\raisebox{+.2cm}{\phantom{\Large{j}}}
$D^{\ast}_s\eta$&0.221 &0.6&0.3&0.4&0.3&0.3&$\sim0.1$\\ 
{\phantom{\Large{l}}}\raisebox{+.2cm}{\phantom{\Large{j}}}
$DK^\ast$&0.561 &2.7&0.71&0.2&0.4&0.2&$0.9\sim 2.5$\\ 
{\phantom{\Large{l}}}\raisebox{+.2cm}{\phantom{\Large{j}}}
$\Gamma_{total}$& 47.6 &67&37&43.2&34&36&$42\sim 60$\\
\hline\hline
\end{tabular*}
\end{table}

\begin{table}
\caption{ Two-body strong decay widths (MeV) of $D_{3}^{\ast}(2760)$. $^3P_0$ model is adopted in Refs.~\cite{Li11}\cite{Chen15}\cite{Lu14}\cite{Yu15} and the chiral quark model is adopted by Ref.~\cite{Zhong10}.}
\vspace{0.2cm}
\setlength{\tabcolsep}{0.1cm}
\centering
\begin{tabular*}{\textwidth}{@{}@{\extracolsep{\fill}}ccccccc}
\hline\hline
Mode&Ours&Ref.~\cite{Chen15}&Ref.~\cite{Yu15}&Ref.~\cite{Li11}&Ref.~\cite{Zhong10}& Ref.~\cite{Lu14}\\
\hline
{\phantom{\Large{l}}}\raisebox{+.2cm}{\phantom{\Large{j}}}
$D\pi$&33.1 &27.9&25.75&31.66&32.5&14.06\\ 
{\phantom{\Large{l}}}\raisebox{+.2cm}{\phantom{\Large{j}}}
$D^{\ast}\pi$&22.0 &15.5&15.67&30.71&20.6&11.09\\ 
{\phantom{\Large{l}}}\raisebox{+.2cm}{\phantom{\Large{j}}}
$D\eta$&0.812 &1.4 &0.99&1.77&2.6&0.77\\ 
{\phantom{\Large{l}}}\raisebox{+.2cm}{\phantom{\Large{j}}}
$D^\ast\eta$&0.254 &0.2 &0.24&0.76&0.7&0.26\\ 
{\phantom{\Large{l}}}\raisebox{+.2cm}{\phantom{\Large{j}}}
$D_sK$&2.30 &1.6&0.70&0.82&2.1&0.22\\
{\phantom{\Large{l}}}\raisebox{+.2cm}{\phantom{\Large{j}}}
$D^\ast_sK$& 0.416&0.2&0.09&0.21&0.3&0.04\\
{\phantom{\Large{l}}}\raisebox{+.2cm}{\phantom{\Large{j}}}
$D\rho$&1.59&0.2&40.16&2.15&0.4&0.66\\ 
{\phantom{\Large{l}}}\raisebox{+.2cm}{\phantom{\Large{j}}}
$D\omega$&0.423&0.1&12.62&0.65&0.1&0.20\\ 
{\phantom{\Large{l}}}\raisebox{+.2cm}{\phantom{\Large{j}}}
$D_1^{\prime}(2430)\pi$& 6.99&1.1&0.065&2.13&5.2&0.37\\
{\phantom{\Large{l}}}\raisebox{+.2cm}{\phantom{\Large{j}}}
$D_1(2420)\pi$& 1.02&0.4&0.024&0.05&1.7&0.03\\
{\phantom{\Large{l}}}\raisebox{+.2cm}{\phantom{\Large{j}}}
$D_2^{\ast}(2460)\pi$& 3.70&1.1&0.17&2.28&1.7&0.62\\
{\phantom{\Large{l}}}\raisebox{+.2cm}{\phantom{\Large{j}}}
$D(2550)\pi$&0.03 &0.0&&&&$5.6\times 10^{-4}$\\
{\phantom{\Large{l}}}\raisebox{+.2cm}{\phantom{\Large{j}}}
$\Gamma_{total}$&72.6&49.7&96.49&73.17&67.9&28.32\\ 
\hline\hline
\end{tabular*}
\end{table}

\begin{table}
\caption{ Two-body strong decay widths (MeV) of the $B_{3}^{\ast}$ state with the mass $6.11$ GeV. The second subrow of the first row is the mass (MeV) of the $B_{3}^\ast$ meson used in different models. The value $a[b]$ represents $a\times10^{-b}$. Refs.~\cite{Sun14}\cite{GMS}\cite{Lu} use the $^3P_0$ model. Ref.~\cite{ZGW14} and Ref.~\cite{Xiao14} use the heavy meson effective theory and chiral quark model, respectively.}
\vspace{0.2cm}
\setlength{\tabcolsep}{0.1cm}
\centering
\begin{tabular*}{\textwidth}{@{}@{\extracolsep{\fill}}cccccccc}
\hline\hline
\multirow{2}{*}{Mode}&\multicolumn{2}{c}{Ours}&Ref.~\cite{Sun14}&Ref.~\cite{ZGW14}&Ref.~\cite{GMS}&Ref.~\cite{Lu}&Ref.~\cite{Xiao14}\\
\cline{2-3}
&6105 &5978&6105&5978&6106&5978&$5950\sim6050$ (5978)\\
 \hline
{\phantom{\Large{l}}}\raisebox{+.2cm}{\phantom{\Large{j}}}
$B\pi$&24.7& 11.7&4.9&37.7&14.4&20.19&\\ 
{\phantom{\Large{l}}}\raisebox{+.2cm}{\phantom{\Large{j}}}
$B^{\ast}\pi$&24.7& 10.3&6.2&31.8&14.2&21.34&\\
{\phantom{\Large{l}}}\raisebox{+.2cm}{\phantom{\Large{j}}}
$B_2^{\ast}\pi$&6.19&$0.546$&0.74&---&0.460&0.31&\\
{\phantom{\Large{l}}}\raisebox{+.2cm}{\phantom{\Large{j}}}
$B_1\pi$&$9.40[2]$&$2.85[3]$&$9.0[2]$&---&0.117&0.15&\\
{\phantom{\Large{l}}}\raisebox{+.2cm}{\phantom{\Large{j}}}
$B_1^\prime\pi$&4.96&0.185&0.17&---&0.0615&0.14&\\
{\phantom{\Large{l}}}\raisebox{+.2cm}{\phantom{\Large{j}}}
$B\eta$&0.43& 0.06&0.21&0.2&0.441&0.31&\\ 
{\phantom{\Large{l}}}\raisebox{+.2cm}{\phantom{\Large{j}}}
$B^{\ast}\eta$&0.31& 0.023&0.20&$<0.1$&0.257&0.14&\\ 
{\phantom{\Large{l}}}\raisebox{+.2cm}{\phantom{\Large{j}}}
$B\rho$&$5.94[2]$&---&$1.8[2]$&---&&&\\ 
{\phantom{\Large{l}}}\raisebox{+.2cm}{\phantom{\Large{j}}}
$B^\ast\rho$&$7.19[2]$&---&1.3&---&&&\\ 
{\phantom{\Large{l}}}\raisebox{+.2cm}{\phantom{\Large{j}}}
$B\omega$&$1.13[2]$&---&$3.7[3]$&---&&&\\ 
{\phantom{\Large{l}}}\raisebox{+.2cm}{\phantom{\Large{j}}}
$B_sK$&1.12 &0.08&$5.4[2]$&0.3&0.366&0.16&\\ 
{\phantom{\Large{l}}}\raisebox{+.2cm}{\phantom{\Large{j}}}
$B_s^\ast K$&0.64& 0.015&$4.5[2]$&$<0.1$&0.197&0.03&\\ 
{\phantom{\Large{l}}}\raisebox{+.2cm}{\phantom{\Large{j}}}
$\Gamma_{total}$&63.3& $22.9$&14&70&31&42.69&$50\sim120$ (60)\\
\hline\hline
\end{tabular*}
\end{table}

\begin{table}
\caption{ Two-body strong decay widths (MeV) of the $B_{s3}^{\ast}$ state with the mass $6.18$ GeV. The second subrow of the first row is the mass (MeV) of the $B_{s3}^\ast$ meson used in different models. The value $a[b]$ represents $a\times10^{-b}$. The $^3P_0$ model is used in Refs.~\cite{Sun14}\cite{GMS}\cite{Lu}, and the chiral quark model is used in Ref.~\cite{Xiao14}.}
\vspace{0.2cm}
\setlength{\tabcolsep}{0.1cm}
\centering
\begin{tabular*}{\textwidth}{@{}@{\extracolsep{\fill}}cccccc}
\hline\hline
\multirow{2}{*}{Mode}&Ours&Ref.~\cite{Sun14}&Ref.~\cite{GMS}&Ref.~\cite{Lu}&Ref.~\cite{Xiao14}\\
&6178&6178&6179&6096&$6050\sim6150$ (6070)
\\ \hline
{\phantom{\Large{l}}}\raisebox{+.2cm}{\phantom{\Large{j}}}
$BK$&$21.1$&5.2&14&23.69&\\ 
{\phantom{\Large{l}}}\raisebox{+.2cm}{\phantom{\Large{j}}}
$B^\ast K$&18.6&5.7&11.4&21.78&\\ 
{\phantom{\Large{l}}}\raisebox{+.2cm}{\phantom{\Large{j}}}
$BK^\ast$&$7.82[5]$&$3.0[5]$&&&\\ 
{\phantom{\Large{l}}}\raisebox{+.2cm}{\phantom{\Large{j}}}
$B_s\eta$&$0.65$&$5.3[2]$&0.522&0.57&\\ 
{\phantom{\Large{l}}}\raisebox{+.2cm}{\phantom{\Large{j}}}
$B_s^{\ast}\eta$&0.41&$4.5[2]$&0.305&0.30&\\
{\phantom{\Large{l}}}\raisebox{+.2cm}{\phantom{\Large{j}}}
$\Gamma_{total}$&40.8&11&26.4&46.33&$25\sim75$ (30)\\
\hline\hline
\end{tabular*}
\end{table}

\begin{table}
\caption{The ratios of decay widths of different channels. }
\vspace{0.2cm}
\setlength{\tabcolsep}{0.01cm}
\centering
\begin{tabular*}{\textwidth}{@{}@{\extracolsep{\fill}}ccccc}
\hline\hline
Mode&$\frac{\Gamma[D_{s3}^\ast(2860)\rightarrow D^{\ast}K]}{\Gamma[D_{s3}^\ast(2860)\rightarrow DK]}$&$\frac{\Gamma[D_{3}^\ast(2760)\rightarrow D^{\ast}\pi]}{\Gamma[D_{3}^\ast(2760)\rightarrow D\pi]}$&$\frac{\Gamma[B_{3}^\ast\rightarrow B^{\ast}\pi]}{\Gamma[B_{3}^\ast\rightarrow B\pi]}$&$\frac{\Gamma[B_{s3}^\ast\rightarrow B^{\ast}K]}{\Gamma[B_{s3}^\ast\rightarrow BK]}$\\ \hline
{\phantom{\Large{l}}}\raisebox{+.2cm}{\phantom{\Large{j}}}
Ours&0.47&0.66&1.0 (0.88)&0.88\\ 
{\phantom{\Large{l}}}\raisebox{+.2cm}{\phantom{\Large{j}}}
Ref.~\cite{Li}&0.75&&&\\ 
{\phantom{\Large{l}}}\raisebox{+.2cm}{\phantom{\Large{j}}}
Ref.~\cite{Chen15}&0.43&0.56&&\\ 
{\phantom{\Large{l}}}\raisebox{+.2cm}{\phantom{\Large{j}}}
Ref.~\cite{God14}&0.60&&&\\ 
{\phantom{\Large{l}}}\raisebox{+.2cm}{\phantom{\Large{j}}}
Ref.~\cite{Li11}&0.75&0.97&&\\ 
{\phantom{\Large{l}}}\raisebox{+.2cm}{\phantom{\Large{j}}}
Ref.~\cite{GMS}&&&0.99&0.81\\ 
{\phantom{\Large{l}}}\raisebox{+.2cm}{\phantom{\Large{j}}}
Ref.~\cite{Xiao14}&0.5&0.65&0.9&0.71\\ 
{\phantom{\Large{l}}}\raisebox{+.2cm}{\phantom{\Large{j}}}
Ref.~\cite{Sun14}&&&1.27&1.10\\ 
{\phantom{\Large{l}}}\raisebox{+.2cm}{\phantom{\Large{j}}}
Exp.~\cite{babar2}&$1.10\pm0.15\pm0.19$&&&\\ 
\hline\hline
\end{tabular*}
\end{table}

\section{Summary}

We have studied OZI-allowed two body strong decays of $3^-$ heavy-light mesons.  The instantaneous Bethe-Salpeter method is applied to get the wave functions of heavy mesons. For $D_{s3}^\ast$ and $D_3^\ast$, the total decay widths are within the experimental error. For $B_3^\ast$ state, we present total and several main decay widths within the mass region $5950\sim6150$ MeV. When $M_{B_3^\ast}=5978$ MeV, our result is much smaller than the central value of the decay width of the new discovered $B(5970)$, while it is still within the experimental errors. So more precise detection is needed. For the $B_{s3}^\ast$ state, there is no candidate in experiments, and our calculations can provide some help for the future study of this particle. Our results also show that the decay widths of $B_3^\ast$ and $B_{s3}^\ast$ depend strongly on the particle mass.

\section*{Acknowledgments}

This work was supported in part by the National Natural Science
Foundation of China (NSFC) under Grant No.~11405037, No.~11575048, No.~11505039, and No.~11405004, and in part by PIRS of HIT No. B201506.



\end{document}